\documentclass[]{spie}  
\usepackage{graphicx,graphics,color,epsfig,amsmath,amsfonts,amssymb,mathrsfs,bm,verbatim,psfrag,times}
\usepackage{flafter,balance}
\usepackage{indentfirst}
\newcommand{\ket}[1]{\left|{#1}\right\rangle}
\newcommand{\bra}[1]{\left\langle{#1}\right|}

\begin{document}

\title{Two coupled Jaynes-Cummings cells}

\author{Peng Xue\supit{a,b}, Zbigniew Ficek\supit{c} and Barry C. Sanders\supit{b}
\skiplinehalf
\supit{a}Department of Physics, Southeast University, Nanjing 211189, P. R. China \\
\supit{b}Institute for Quantum Information Science, University of
Calgary, Alberta T2N 1N4, Canada\\
\supit{c}The National Centre for Mathematics and Physics, KACST,
Riyadh 11442, Saudi Arabia}

\authorinfo{Further author information: (Send correspondence to Peng Xue)
\\Peng Xue.: E-mail: gnep.eux@gmail}

 \maketitle

\begin{abstract}
We develop a theoretical framework to evaluate the energy spectrum, stationary states, and dielectric susceptibility of two Jaynes-Cummings systems coupled together by the overlap of their respective longitudinal field modes, and we solve and characterize the combined system for the case that the two atoms and two cavities share a single quantum of energy.
\end{abstract}

\section{introduction}

The Jaynes-Cummings (JC) system~\cite{Pau63,JC63} comprises a two-level atom (2LA) coupled to a single optical resonator mode.
Although Jaynes and Cummings emphasized the robustness of the semiclassical description as compared to the fully quantum model,
manifestations of quantum field effects are now ubiquitous~\cite{ENS80,CKS96,SCW97,BSM+96,FGB+08}
leading to optimism that JC systems will soon be coupled together in one- or more-dimensional lattices~\cite{CZKM97}
thereby yielding novel condensed-matter phenomena~\cite{HBP06,GTCH06,RF07,ASB07}. 
The first step to achieving coupled JC systems is to create and study a double-JC (DJC) system.

We develop a theoretical framework for the DJC system by calculating its stationary states, energy spectrum and dielectric susceptibility,
and we show that this quadripartite system comprising two atoms and two field modes has fascinating features.
The system is effectively characterized by two independent parameters:
$g$ for the coupling rate between the resonator and the single atom
and~$\kappa$ for the coherent photon hopping rate between the two resonators and proportional to the overlap of the two resonator field modes.
By varying~$g$ and~$\kappa$, quite different features emerge from the DJC system.
Although we are interested in general properties of this system,
our focus here is specifically on the case that the the DJC system shares precisely one quantum of energy.

\section{Double Jaynes-Cummings system}

The JC system has an atomic dipole with frequency~$\omega_\text{a}$ coupled
to a cavity with frequency~$\omega_\text{c}$ via the atom's electric
dipole between ground state~$\ket{\text{g}}$ and~$\ket{\text{e}}$.
The dipole coupling frequency is~$g$, and the atom-cavity decoupling
frequency is $\Delta =\omega_\text{a}-\omega_\text{c}$.
In the rotating-wave approximation, the JC Hamiltonian is $(\hbar\equiv 1)$:
\begin{equation}
	\hat{H}^\text{JC} =\omega_\text{c}\left(\hat{a}^\dagger\hat{a}+\frac{1}{2}\right)+\frac{1}{2}\omega_\text{a}\hat{\sigma}_z
		+g\left(\hat{a}^\dagger\hat{\sigma}_-+\hat{a}\hat{\sigma}_+\right) ,
\end{equation}
where $\hat{a}^{\dagger}$ and $\hat{a}$ are creation and
annihilation operators for the cavity field and $\hat{\sigma}_{\pm}$
and $\hat{\sigma}_z$ with $\text{spec}(\hat\sigma_z)=\pm1$ are
spin operators for the atoms.
The energy spectrum of the JC system is
\begin{equation}
	\omega^{\text{JC}(0)}=-\Delta/2,\,
	\omega^{\text{JC}(n)}_\pm=n\omega_c\pm\sqrt{ng^2+\Delta^2/4}
\end{equation}
for the ground state and the $n^{\rm th}$ JC doublet, respectively.
The stationary states are the ground state singlet~$\ket{0\text{g}}$ and the excited state doublets
$\ket{\pm}_n$ given by
\begin{equation}
\ket{+}_n\pm\text{i}\ket{-}_n
=\text{e}^{\mp\text{i}\theta_n}\ket{n,\text{g}}+\text{e}^{\pm\text{i}\theta_n}\ket{n-1,\text{e}}
,\theta_n =\tan^{-1}(2g\sqrt{n}/\Delta)/2.
\end{equation}

\begin{figure}
\label{fig:scheme}
	\begin{center}
		\includegraphics[width=0.5\columnwidth]{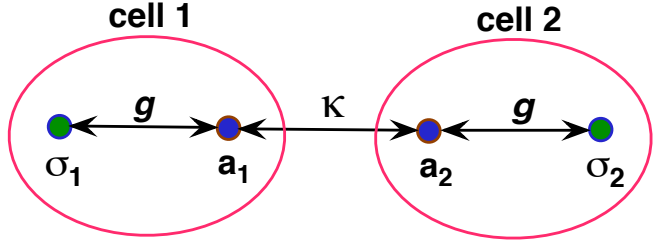}
	\end{center}
	\caption{Scheme for the system comprising two resonators each containing a two-level atom $\sigma_i$ and a bosonic mode $a_i$ for $i=1,2$.}
\end{figure}

Two neighboring cells (JC cavities) have overlapping evanescent mode functions resulting in an intercavity hopping rate~$\kappa$
depicted schematically in Fig.~\ref{fig:scheme}.
For~$\hat{a}_i$ and~$\hat{\sigma}_i$ the field annihilation operator
and atomic electron energy lowering operator for the $i^\text{th}$
JC system ($i=1,2$), the DJC Hamiltonian is
\begin{equation}
	\hat{H}=\hat{H}_1^\text{JC}+\hat{H}_2^\text{JC}-\kappa(\hat{a}_1^\dagger\hat{a}_2+\hat{a}_1\hat{a}_2^\dagger)=\bigoplus_\nu\hat{H}^{(\nu)}
\end{equation}
with~$\nu$ the total number
of quanta shared between the two atoms and the two cavity modes.
Thus, the pure-state Hilbert space~$\mathscr{H}$ is a union of subspaces~$\mathscr{H}^{(\nu)}$ with definite overall particle number~$\nu$.

Although we are interested in characterizing this system for all~$\nu$,
our focus in this paper is solely on studying the $\nu=1$, which is intriguing in its own right.
The $\nu=1$ case provides an enticing simplification:
the field modes can be treated as two-level systems (known as `qubits' in quantum information parlance).
As the system has atoms coupled only to the field modes and the field modes coupled to each other,
for $\nu=1$, the system corresponds to a chain of four qubits as shown in Fig.~\ref{fig:scheme} with at most one qubit in the upper state.

The spectrum and stationary states for the cases $\nu=0,1$ can be solved in closed form.
Let $\mathscr{B}^{(\nu)}=\{\ket{n_1,\text{c}_1,n_2,\text{c}_2}\}$
be a basis for~$\mathscr{H}^{(\nu)}$ with~$n_i$ denoting the
number of photons in the $i^{\rm th}$ mode and $\text{c}_i=(\text{e,g})$
the state of the $i^{\rm th}$ atom. 
For the trivial case of no excitation
in the system $\nu=0$, we have $\mathscr{B}^{(0)}=\{\ket{0{\text
g}0\text{g}}\}$.
For one excitation $(\nu=1)$ there are four basis states
\begin{equation}
	\mathscr{B}^{(1)}=\{\ket{0\text{e}0\text{g}},\ket{1\text{g}0\text{g}},\ket{0\text{g}1\text{g}},\ket{0\text{g}0\text{e}}\},
\end{equation}
In this basis
\begin{equation}
    \hat{H}^{(1)}=\begin{pmatrix}
    0 & g & 0 & 0 \\
    g & -\Delta & -\kappa & 0 \\
    0 & -\kappa & -\Delta & g \\
    0 & 0 & g & 0
    \end{pmatrix}
    +\omega_\text{c}I ,
\end{equation}
with~$I$ the $4 \times 4$ identity matrix.
The Hamiltonian matrix is not diagonal due to the presence of the coupling~$\kappa$.
For independent cells, $\kappa=0$, and then the
$\hat{H}^{(1)}$ matrix becomes a block diagonal with two $2\times 2$
matrices each corresponding to a single cell.

\begin{figure}
\label{fig:spectrum}
	\begin{center}
		\includegraphics[height=5.5cm]{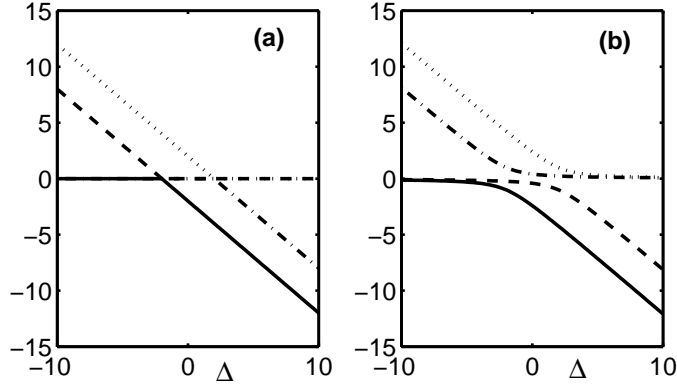}
	\end{center}
	\caption[example] {The $\nu=1$ energy spectrum vs detuning $\Delta$ illustrating avoided crossing effect to occur only for $g\neq 0$ and at $\Delta=\pm\kappa$: (a)~$g=0, \kappa=2$ and (b)~$g=1, \kappa=2$.}
\end{figure}

By diagonalizing the Hamiltonians $\hat{H}^{(\nu)}$ for $\nu=0,1$,
the corresponding energy spectra are determined to be
\begin{equation}
	\omega^{(0)}=-\Delta=2\omega^{\text{JC}(0)} , \,
	\omega^{(1)}_{\epsilon\pm}
		=\omega_c-\frac{1}{2}(\Delta+\epsilon\kappa)\pm\sqrt{g^2+\frac{1}{4}\left(\Delta+\epsilon\kappa\right)^2}
\end{equation}
with $\epsilon=\pm$. The spectrum $\omega^{(0)}$ is composed of a
single nondegenerate state that is independent of $g$ and $\kappa$,
whereas the spectrum $\omega^{(1)}$ is composed of doubly degenerate
pairs for $\kappa=0$ and is non-degenerate for $\kappa\neq 0$.
The coherent photon hopping effect $(\kappa\neq 0)$ does
not lead to avoided crossings of the energies.
Rather coherent hopping shifts the crossing point from $\Delta =0$ to $\Delta =\pm \kappa$.
Only the atom-cavity coupling rate~$g$ results in avoided crossings.
This avoided-crossing phenomenon for~$g$ and not for~$\kappa$ is
shown in Fig.~\ref{fig:spectrum}, which shows the eigenvalues
$\omega^{(1)}_{\epsilon\pm}$ as a function of $\Delta$ for
$\kappa\neq 0, g=0$ (left frame) and $\kappa\neq 0, g\neq 0$ (right frame). 
Evidently the energies cross at $\Delta=\pm\kappa$ for~$g=0$, and $g\neq 0$ leads to avoided crossing.

For zero coupling $\kappa=0$, the first-level spectral term reduces to two degenerate eigenvalues
\begin{equation}
	\omega^{(1)}_{\epsilon\pm}|_{\kappa=0}=\omega^{\text{JC}(0)}+\omega_\pm^{\text{JC}(1)},
\end{equation}
with~$\epsilon$ irrelevant.
Thus, the spectral values for
$\kappa=0$ can be understood as sums of spectral values for each of
the two isolated JC systems.

Now consider the $\nu=1$ spectrum for the limited case of $g=0$ and
let $\Delta=0$. In this case, $\omega^{(0)}=0$, and
$\omega^{(1)}_{\epsilon\pm}|_{g=0=\Delta}$ values are $\omega_c$
(doubly-degenerate) and $\omega_c\pm\kappa$.
These spectra correspond to the spectra for coupled harmonic oscillators as expected. Thus,
strong inter-cavity mode coupling (large~$\kappa$) compared to
atom-cavity coupling ($g$) is expected to make the DJC system behave
nearly like coupled harmonic oscillators with a perturbation due to
atom-cavity coupling.

In the other limit we consider small~$\kappa$ with strong
coupling~$g$.
For $\nu=1$ we have the spectral values
\begin{equation}
	\omega^{(1)}_{\epsilon\pm}\approx\omega_c\pm g-\frac{\epsilon\kappa}{2}\pm\frac{\kappa^2}{8g}.
\end{equation}
The shift $\pm g$ is due to vacuum Rabi splitting, and $\epsilon\kappa/2$ is normal-mode splitting due to inter-cavity coupling. 
The next-order shift $\pm\kappa^2/8g$ is
analogous to the ac Stark shift: the field of one cavity induces
frequency pulling on the other cavity, which is detuned by~$g$ due
to vacuum Rabi splitting. 
Thus, the strong-$g$, weak-$\kappa$ limit is equivalent to a weakly driven
strong-coupling JC model.

We now focus on the eigenstates of the system for $\nu=0,1$. The
$\nu=0$ case is trivial, composed of a singlet
$\ket{0\text{g}0\text{g}}$ whose energy is independent of~$g$
and~$\kappa$. The eigenstates of the $\nu=1$ Hamiltonian are
\begin{align}
\label{eq:states}
	\ket{\pm,r_{\epsilon}}=u_{\epsilon}^\pm\!\left(\ket{1\text{g}0\text{g}}\!-\!\epsilon\ket{0\text{g}1\text{g}}\right)\!+\!w_{\epsilon}^\pm\!\left(\ket{0\text{e}0\text{g}}\!-\epsilon\!\ket{0\text{g}0\text{e}}\right)
,
\end{align}
in which
\begin{align}
u^\pm_\epsilon
=\frac{-r_\epsilon\pm\sqrt{1+r_{\epsilon}^2}}{\sqrt{2+2\left(r_{\epsilon}\mp\sqrt{1+r_{\epsilon}^2}\,\right)^2}}
,\quad w^\pm_\epsilon
=\frac{1}{\sqrt{2+2\left(r_{\epsilon}\mp\sqrt{1+r_{\epsilon}^2}\,\right)^2}}
,\label{uw}
\end{align}
with
\begin{align}
\label{r}
r_\epsilon = \frac{\Delta+\epsilon\kappa}{2g}.
\end{align}
The states $\ket{\pm,r_{\epsilon}}$ in Eq.~(\ref{eq:states}) are like W~states, i.e., a
superposition of one excitation in each of the four degrees of freedom.

In general, the states~(\ref{eq:states}) are non-maximally entangled states
with unequal weighting of the field and atomic states in the superposition. 
The states are maximally entangled for $u^\pm_\epsilon=w^\pm_\epsilon=1/2$.
A close look at Eq.~(\ref{uw})
reveals that this could happen only for $r_{\epsilon} =0$, which,
according to (\ref{r}), only takes place for $\Delta=\pm \kappa$.
Thus, maximally entangled four-qubits states can be created at the
thresholds $\Delta=-\kappa$ and $\Delta=\kappa$.

\section{Linear susceptibility}

The spectral properties and stationary states reveal the nature of
the DJC system, but ultimately these features need to be observed
experimentally. One way to observe these properties is to measure
the dielectric susceptibility for a probe field directed through
both cavities sequentially with measurement of the output field. The
susceptibility is especially important for characterizing the DJC
system for two reasons: susceptibility is
experimentally meaningful on a macroscopic scale (i.e., without
needing to manipulate individual elements of the system such as a
single atom or cavity), and the susceptibility would provide a direct
signature of a quantum phase transitions in the JC lattice case.

The connection between susceptibility and spectrum is as follows.
Given that the system is prepared in some sector of  $\nu$ excitations,
the probe field will excite the system only
if the field frequency is close to resonant with a transition from
this energy state to a higher energy state in the system. The susceptibility is proportional 
to the probability of the transitions from $\nu=1$ states to the ground state.
For the system prepared in the sector~$\nu=1$, the susceptibility is given by
\begin{align}
    \bm{\chi}^{(1)}_0\left(\omega_\text{p}\right) =\sum_{i=\pm}\sum_{\epsilon=\pm} \Gamma_{i,\epsilon}
           \left(\omega_{\epsilon i}^{(1)}-\omega_{p}-\text{i}\gamma_\text{a}\right)^{-1} ,\label{sus}
\end{align}
where $\gamma_\text{a}$ is a small parameter describing a finite width of the transitions and
\begin{align}
\Gamma_{i,\epsilon}=\gamma\left|\bra{i,r_{\epsilon}}\left(\hat{\sigma}_1^{+}+\hat{\sigma}_2^{+}\right)\ket{0}\right|^2
+\gamma_\text{c}\left|\bra{i,r_{\epsilon}}\left(\hat{a}_1^{\dag}+\hat{a}_2^{\dag}\right)\ket{0}\right|^2
\end{align}
is the total probability of the transitions from the energy states $\ket{\pm,r_{\epsilon}}$ to the ground
state~$\ket{0}\equiv\ket{0\text{g}0\text{g}}$. 
The total probability of the transitions $\Gamma_{i,\epsilon}$ is a sum of the
squares of the absolute values of the amplitudes of transitions from
the energy states $\ket{\pm,r_{\epsilon}}$ to the ground
state~$\ket{0}$ caused by spontaneous emission from the atoms,
occurring with the rate~$\gamma$, and by damping of the modes of the
JC cavities with the rate $\gamma_\text{c}$.

The atomic dipole operators and the field operators have matrix elements
\begin{equation}
	\bra{\pm,r_{\epsilon}}\left(\hat{\sigma}_+^1+\hat{\sigma}_+^2\right)\ket{0} = (1-\epsilon)w_{\epsilon}^\pm,
	\bra{\pm,r_{\epsilon}}\left(\hat{a}_1^{\dag} +\hat{a}_2^{\dag}\right)\ket{0}= (1-\epsilon)u_{\epsilon}^\pm,
\end{equation}
which cause transitions from the $\nu=1$ states
$\ket{\pm,r_{\epsilon}}$ to the ground state~$\ket{0}$ to occur with
probabilities
\begin{align}
\Gamma_{\pm,\epsilon}=
\left(1-\epsilon\right)^2\left(\gamma\left|w_{\epsilon}^\pm\right|^2
+\gamma_\text{c}\left|u_{\epsilon}^\pm\right|^2\right).\label{gam}
\end{align}
Clearly the transition rates from states with $\epsilon=+1$ are zero so that the states are dark (non-radiative) states irrespective of the spontaneous emission, cavity damping, and whether the states are maximally entangled or not.
This property contrasts with the two-qubit case for which a dark state can be created between the qubits only if the qubits are identical~\cite{FT02}.

In our case of four qubits, the atoms are degenerate in frequency and the field modes are also degenerate in frequency, but these frequencies differ from each other.
Of course, the states $\epsilon=+1$ would not be dark states when we unbalance the symmetry between the atoms and/or the cavity modes, i.e. when either the atoms or the field modes are not degenerate in frequency.
The symmetry could also be broken by allowing the atoms and/or the cavity modes to be damped with different rates.
If the atoms are damped with different rates, say $\gamma_1$ and $\gamma_2$, and the cavity modes are damped with rates $\gamma_{c1}$ and $\gamma_{c2}$, respectively, then the transitions between the states $\ket{\pm,r_{\epsilon}}$ and~$\ket{0}$ occur with probabilities
\begin{align}
\label{e1}
	\Gamma_{\pm,\epsilon}= \left(\sqrt{\gamma_1}-\epsilon\sqrt{\gamma_2}\right)^2\left|w_{\epsilon}^\pm\right|^2
		+\left(\sqrt{\gamma_{c1}}-\epsilon\sqrt{\gamma_{c2}}\right)^2\left|u_{\epsilon}^\pm\right|^2.
\end{align}
Clearly, the transition probabilities are different from zero
irrespective of $\epsilon$. In this case, the absorption spectrum is
composed of four peaks of different magnitudes.

For the case that $\gamma_1=\gamma_{c1}$ and $\gamma_2=\gamma_{c2}$, the probabilities (\ref{e1}) reduce to
\begin{equation}
	\Gamma_{\pm,\epsilon}= \left(\sqrt{\gamma_1}-\epsilon\sqrt{\gamma_2}\right)^2  .\label{e1a}
\end{equation}
In this case, transitions from the two states corresponding to $\epsilon=+1$ occur with the same probability $(\sqrt{\gamma_1}-\sqrt{\gamma_2})^2 $. Similarly, transitions from the states corresponding to $\epsilon=-1$ occur with probability $(\sqrt{\gamma_1}+\sqrt{\gamma_2})^2 $. Thus, the resulting absorption spectrum of a probe field is expected to be symmetric regardless of whether the states $\ket{\pm,r_{\epsilon}}$ are maximally entangled or~not.
Thus, if the atom and the cavity mode of cell~1 are damped with the same rates and also the atom and the cavity mode of the cell~2 are damped with the same rates, which may or may not equal the damping rates of cell~1, the transition probabilities are independent of whether the states $\ket{\pm,r_{\epsilon}}$ are maximally entangled or not.

However, in the case of $\gamma\neq\gamma_\text{c}$, the spectrum could be symmetric only if the states $\ket{\pm,r_{\epsilon}}$ are maximally entangled. Otherwise, the spectrum is asymmetric.
From Eqs.~(\ref{sus}) and (\ref{gam}), the imaginary part of the susceptibility (absorption spectrum) is a sum of two Lorentzians:
\begin{align}
    \text{Im}\left[\bm{\chi}^{(1)}_0\left(\omega_\text{p}\right)\right] =\frac{\gamma_\text{a}\Gamma_{+,-}}
           {\left(\omega_{-+}^{(1)}-\omega_{p}\right)^2 +\gamma_\text{a}^2} + \frac{\gamma_\text{a}\Gamma_{-,-}}
           {\left(\omega_{--}^{(1)}-\omega_{p}\right)^2 +\gamma_\text{a}^2}  .
\end{align}
The symmetry of the spectrum depends on the ratio between $\Gamma_{+,-}$ and $\Gamma_{-,-}$,
and the spectrum could be symmetric only for $\Gamma_{+,-}=\Gamma_{-,-}$.
From Eq.~(\ref{gam}) we see that $\Gamma_{+,-}=\Gamma_{-,-}$ holds only if $\left|u_{\epsilon}^\pm\right|^2=\left|w_{\epsilon}^\pm\right|^2=1/4$.
According to Eq.~(\ref{gam}), it happens only when the states $\ket{\pm,r_-}$ are maximally entangled. Hence, an observation of the symmetric absorption spectrum when $\gamma\neq \gamma_\text{c}$ could be regarded as an indication of the presence of maximally entangled states in the system.
\begin{figure}
	\begin{center}
	\includegraphics[width=0.6\columnwidth]{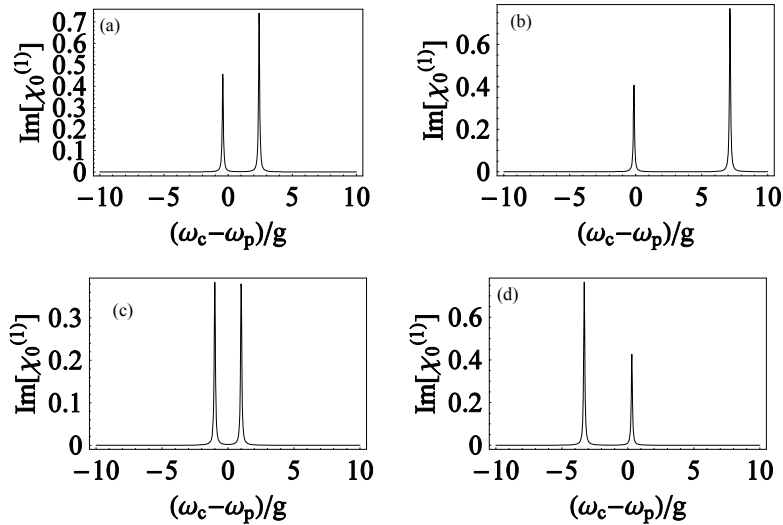}
	\end{center}
\caption{Absorption spectra~$\text{Im}\left[\chi^{(1)}_{0}\left(\omega_\text{p}\right)\right]$
of two coupled JC cells plotted as a function of $(\omega_{\text{c}}-\omega_{\text{p}})/g$ for $\gamma_\text{a}/g=0.05$,
$\gamma/g=0.01$, $\gamma_c/g=0.02$, $\kappa/g=2$ and different $\Delta$: 
(a)~$\Delta/g=0$, (b)~$\Delta/g=-5$, (c)~$\Delta/g=2$ and
(d)~$\Delta/g=5$.} \label{fig4}
\end{figure}
Figure~\ref{fig4} shows the absorption spectrum for $\gamma\neq\gamma_\text{c}$, $\kappa=2g$, and different $\Delta$.
We see that as long as $\Delta\neq \kappa$, the spectrum is composed of two peaks of unequal amplitudes. The spectrum becomes symmetric at $\Delta=\kappa$. In this case
the states $\ket{\pm,r_-}$ are maximally entangled states. Thus, the
symmetry of the spectrum can be regarded as an indication of the
presence of maximally entangled states.
\begin{figure}
	\begin{center}
	\includegraphics[width=0.7\columnwidth]{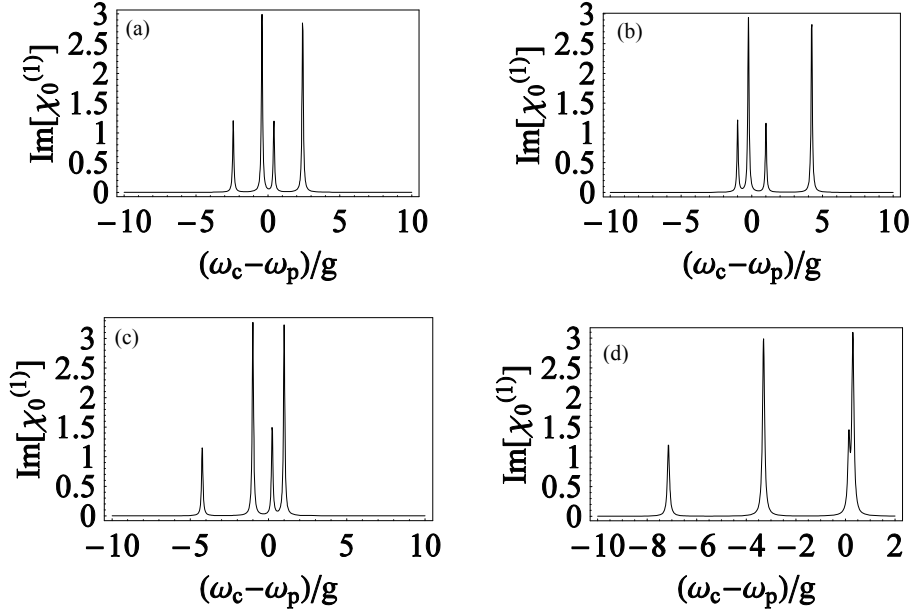}
	\end{center}
\caption{Absorption spectra~$\text{Im}\left[\chi^{(1)}_{0}\left(\omega_\text{p}\right)\right]$ 
of two coupled JC cells plotted as a function of $(\omega_{\text{c}}-\omega_{\text{p}})/g$ for $\gamma_\text{a}/g=0.05$,
$(\gamma_1/g,\gamma_2/g) =(0.01,0.2)$, $(\gamma_{c1}/g,\gamma_{c2}/g)=(0.2,0.01)$, $\kappa/g=2$ and different $\Delta$: 
(a)~$\Delta/g=0$, (b)~$\Delta/g=-2$, (c)~$\Delta/g=2$ and
(d)~$\Delta/g=5$.} 
\label{fig5}
\end{figure}
Figure~\ref{fig5} shows the absorption spectra for $\gamma_1\neq\gamma_2$
and $\gamma_{c1}\neq\gamma_{c2}$ with  $(\gamma_1,\gamma_2)\neq (\gamma_{c1},\gamma_{c2})$. In this
case the spectrum is composed of four peaks and is always asymmetric.
However, one can notice from the figure that, at $\Delta=-\kappa$, two peaks corresponding to transitions from the states $\ket{\pm,r_+}$ have equal amplitudes and are symmetrically located about $(\omega_{\text{c}}-\omega_{\text{p}})/g =0$.
Similarly, at $\Delta=\kappa$, two other peaks that correspond to transitions from the states $\ket{\pm,r_+}$ now have equal amplitudes and are symmetrically located about $(\omega_{\text{c}}-\omega_{\text{p}})/g =0$. Again, equal amplitudes of the peaks indicate that the states which the transitions correspond to are maximally entangled states.

The presence of the threshold values for $\Delta$ at which maximally
entangled four-qubit states are created could be predicted from the
structure of the Hamiltonian of the system.
Instead, working in terms
of the two coupled JC systems, we can introduce symmetric and
antisymmetric combinations of the atomic and field operators
(collective modes)
\begin{align}
	\hat{A}_1 &= \frac{1}{\sqrt{2}}\left(\hat{a}_1+\hat{a}_2\right) ,\quad \hat{A}_2 =\frac{1}{\sqrt{2}}\left(\hat{a}_1-\hat{a}_2\right) ,\nonumber\\
	\hat{S}_1 &=\frac{1}{\sqrt{2}}\left(\hat{\sigma}^-_1+\hat{\sigma}^-_2\right) ,\quad
	\hat{S}_2 =\frac{1}{\sqrt{2}}\left(\hat{\sigma}^-_1-\hat{\sigma}^-_2\right),
\end{align}
and find that, for any $\nu$, the Hamiltonian of the system can be written as
\begin{equation}
	\hat{H} = (\omega_\text{c} +\kappa)\hat{A}_1^{\dag}\hat{A}_1 +(\omega_\text{c} -\kappa)\hat{A}_2^{\dag}\hat{A}_2
		+\frac{1}{2}\omega_{\text{a}}\hat{S}_z +g\left(\hat{A}_1^{\dag}\hat{S}_1+\hat{A}_2^{\dag}\hat{S}_2 + \text{H.c.}\right).\label{hc}
\end{equation}
Thus, the DJC system is equivalent to two independent and
non-degenerate collective systems corresponding to symmetric and
antisymmetric combinations of the modes. The collective bosonic
modes are coupled to the collective atomic system with the same
coupling strength $g$.

Hamiltonian (\ref{hc}) can be written as
\begin{equation}
\label{hc1}
	\hat{H} = \omega_\text{a}\left(\hat{A}_1^{\dag}\hat{A}_1 + \hat{A}_2^{\dag}\hat{A}_2 +\frac{1}{2}\hat{S}_z\right)
		-(\Delta -\kappa)\hat{A}_1^{\dag}\hat{A}_1 - (\Delta+\kappa)\hat{A}_2^{\dag}\hat{A}_2 +g\left(\hat{A}_1^{\dag}\hat{S}_1+\hat{A}_2^{\dag}\hat{S}_2 + \text{H.c.}\right).
\end{equation}
with the first part representing the average free energy of
the field modes and the atoms, the second part representing the
shift of the energies of the superposition modes from the average
energy, and the last part representing the interaction of the field
modes with the collective atomic systems.

Notice that, at $\Delta=\kappa$, the systems $A_1$ and $S_1$ are resonant, so that the coupled two-qubit system can then be maximally
entangled. Similarly, at $\Delta=-\kappa$, the systems $A_2$ and $S_2$ are resonant and therefore can be maximally entangled at that frequency.
Clearly Eq.~(\ref{eq:states}) implies that, at $\Delta=-\kappa$, the avoided crossing occurs between states
$\ket{\pm,r_+}$ that involve antisymmetric combinations of the atomic and field states, whereas, at $\Delta=\kappa$, the avoided
crossing occurs between states~$\ket{\pm,r_-}$.
Thus maximally entangled states can be created in the system for $\Delta=\kappa$ and~$\Delta=-\kappa$.

\section{Conclusions}

We have constructed a framework for calculating the energy spectrum, stationary states, and dielectric susceptibility of two Jaynes-Cummings systems coupled
together by the overlap of their respective longitudinal field modes and solved it for $\nu=0$ and $\nu=1$ excitations of the system, which
can be understood in terms of four coupled qubits.
 For weak coupling, the pair of
systems is similar to a single Jaynes-Cummings system undergoing an
AC Stark effect, and for strong coupling the behavior is similar to
two coupled harmonic oscillators. For moderate coupling strengths,
the pair of atoms and the pair of field modes can be highly entangled
states, and, where the spectrum exhibits avoided crossings as a
function of the detuning, the atoms and fields are found in maximally entangled 
four-qubit W-like states. We also show the susceptibility and
the absorption of the system that explore the entangled features of the system.

\acknowledgments

This work has been supported by NSERC, MITACS, CIFAR, QuantumWorks,
iCORE and NSFC 11004029. BCS is supported by a CIFAR Fellowship.

\bibliography{report2}   
\bibliographystyle{spiebib}   
\end{document}